%% file: aph.tex
\documentclass[twocolumn]{aastex63}
\usepackage{color}

\shorttitle{Where's the Dust?: Extragalactic AME in NGC\,4725\,B}
\shortauthors{MURPHY ET AL.}
\accepted{to ApJ Letters; November 5, 2020}

\begin{document}


\title{Where's the Dust?: The Deepening Anomaly of Microwave Emission in NGC\,4725\,B}

\author{E.J.\,Murphy}
\email{emurphy@nrao.edu}
\affiliation{National Radio Astronomy Observatory, 520 Edgemont Road, Charlottesville, VA 22903, USA}

\author{B.S.\,Hensley}
\affiliation{Department of Astrophysical Sciences, Princeton University, Princeton, NJ 08544, USA; Spitzer Fellow}

\author{S.T.\,Linden}
\affiliation{Department of Astronomy, University of Virginia, 530 McCormick Road, Charlottesville, VA 22904, USA}

\author{B.T.\,Draine}
\affiliation{Department of Astrophysical Sciences, Princeton University, Princeton, NJ 08544, USA}

\author{D.\,Dong}
\affiliation{California Institute of Technology, MC 100-22, Pasadena, CA 91125, USA}

\author{E.\,Momjian}
\affiliation{National Radio Astronomy Observatory, P.O. Box O, 1003 Lopezville Road, Socorro, NM 87801, USA}

\author{G.\,Helou}
\affiliation{Infrared Processing and Analysis Center, California Institute of Technology, MC 220-6, Pasadena CA, 91125, USA}

\author{A.S.\,Evans}
\affiliation{Department of Astronomy, University of Virginia, 3530 McCormick Road,Charlottesville, VA 22904, USA}
\affiliation{National Radio Astronomy Observatory, 520 Edgemont Road, Charlottesville, VA 22903, USA}


\begin{abstract}
We present new Atacama Large Millimeter Array (ALMA) observations towards NGC\,4725\,B, a discrete, compact, optically-faint region within the star-forming disk of the nearby galaxy NGC\,4725 that exhibits strong anomalous microwave emission (AME).  
These new ALMA data include continuum observations centered at 92, 133, 203, and 221\,GHz accompanied by spectral observations of the $^{12}$CO ($J=2\rightarrow1$) line.    
NGC\,4725\,B is detected in the continuum at all frequencies, although the detection at 203\,GHz is marginal.  
While molecular gas is not detected at the exact location of NGC\,4725\,B, there is molecular gas in the immediate vicinity (i.e., $\lesssim 100$\,pc) along with associated diffuse 8\,$\mu$m emission.  
When combined with existing Very Large Array continuum data at 1.5, 3, 5.5, 9, 14, 22, 33, and 44\,GHz, the spectrum is best fit by a combination of AME, synchrotron, and free-free emission that is free-free absorbed below $\sim6$\,GHz.  Given the strength of the AME, there is surprisingly no indication of millimeter dust emission associated with NGC\,4725\,B on $\lesssim$6\arcsec~spatial scales at the sensitivity of the ALMA interferometric data.  
Based on the properties of the nearest molecular gas complex and the inferred star formation rate, NGC\,4725\,B is consistent with being an extremely young ($\sim $3--5\,Myr) massive ($\lesssim 10^{5}\,M_{\odot}$) cluster that is undergoing active cluster feedback. 
However, the lack of millimeter thermal dust emission is difficult to reconcile with a spinning dust origin of the 30\,GHz emission.  
On the other hand, modeling NGC\,4725\,B as a new class of background radio galaxy is also unsatisfactory.  



\end{abstract}

\section{Introduction}
It has been over 20 years since the initial discovery of excess $\sim30$\,GHz ($\sim 1\,$cm), dust-correlated emission in cosmic microwave background experiments \citep[e.g.,][]{ak96b,eml97}, yet our current understanding of the physical mechanism and environmental conditions powering anomalous microwave emission (AME) remains highly incomplete \citep[for a recent review, see][]{AME18}.  
The leading explanation for AME is ``spinning dust," in which rapidly rotating ultrasmall (radius $a \lesssim 1$\,nm) grains with a non-zero electric dipole moment produce a peaked microwave emission spectrum in the frequency range spanning $\approx 10 - 60$\,GHz \citep{wce57,dl98b,ahd09,hoang11,hd17}.  

To date, AME has been observed in a wide variety of astrophysical environments, indicating that it is truly a ubiquitous emission component in the interstellar medium (ISM) of galaxies, and that a complete model may significantly impact our understanding of the astrophysics of the ISM.  
For instance, investigations within the Galaxy have uncovered strong detections of AME in molecular clouds \citep[e.g.,][]{plsd11} along with discrete locations adjacent to larger H\,{\sc ii} complexes \citep[e.g.,][]{cd09,tibbs11}, consistent with the theoretical expectation that it is a common emission component in galaxies. 
Most recently, detection of AME has been reported in proto-planetary disks around Herbig Ae stars \citep{greaves18}.
Although seemingly a common phenomenon from Galactic sources, AME has been observed in surprisingly few extragalactic sources, with only NGC\,6946 \citep{ejm10,as10,bh15}, NGC\,4725 \citep{ejm18c}, and M\,31 \citep{Battistelli_2019} having firm detections. 
This dearth of extragalactic detections likely arises from the lack of deep, wide-field observations of nearby extragalactic sources at $\sim30$\,GHz, the insensitivity of interferometers to large angular scale emission, and the need for multi-frequency data \citep{bh15,stl20}.

In this letter, we revisit the recent detection of AME from an optically-faint region within the star-forming galaxy NGC\,4725 from 33\,GHz Karl G. Jansky Very Large Array (VLA) observations carried out as part of the Star Formation in Radio Survey \citep[SFRS;][]{ejm18a, stl20}.  
This source, NGC\,4725\,B, was studied by \citet{ejm18c}, who concluded that it was consistent with a highly-embedded ($A_V > 5$\,mag) nascent star-forming region, in which young ($\lesssim 3$\,Myr) massive stars are still enshrouded by their natal cocoons of gas and dust, lacking enough supernovae to produce measurable synchrotron emission.  
In light of our new Atacama Large Millimeter Array (ALMA) continuum and molecular gas observations of NGC\,4725\,B, along with additional archival VLA data, this scenario remains largely viable. However, the ALMA data raise a new and unexpected question regarding the origin of the AME---how can such a strong 30\,GHz excess be produced without an accompanying millimeter dust continuum?

This letter is organized as follows:
The data and analysis procedures are presented in Section~\ref{sec:data}. 
The results are then presented in Section~\ref{sec:results}.  
A brief discussion of our findings and main conclusions is given in Section~\ref{sec:summary}.

\input{obsdata}

\section{Data and Analysis}\label{sec:data}
We refer the reader to \citet{ejm18c} for detailed information on NGC\,4725 and NGC\,4725\,B, along with the associated data and analysis procedures presented in that paper.   
In this letter we simply summarize key details of that work and present the new ALMA and archival VLA data, associated analysis procedures, and discuss how these new data affect the previous interpretation.

NGC\,4725 is a nearby \citep[$d_{L} = 11.9$\,Mpc;][]{hkp01} barred ringed spiral galaxy that hosts an AGN (Sy2) nucleus \citep{jm10}.  
Located at J2000 $\alpha =12^\mathrm{h}50^\mathrm{m}28\fs48, \delta =+25\degr30\arcmin22\farcs5$ (at a galactocentric radius of $\approx$1.9\,kpc), NGC\,4725\,B was first detected at 33\,GHz as part of SFRS \cite[][]{ejm18a}, comprising nuclear and extranuclear star-forming regions in 56 nearby galaxies ($d_{L} < 30$\,Mpc) observed as part of the {\it Spitzer} Infrared Nearby Galaxies Survey \citep[SINGS;][]{rck03} and Key Insights on Nearby Galaxies: a Far-Infrared Survey with {\it Herschel} \citep[KINGFISH;][]{kf11} legacy programs.

\begin{figure*}[t!]
\epsscale{1.}
\plotone{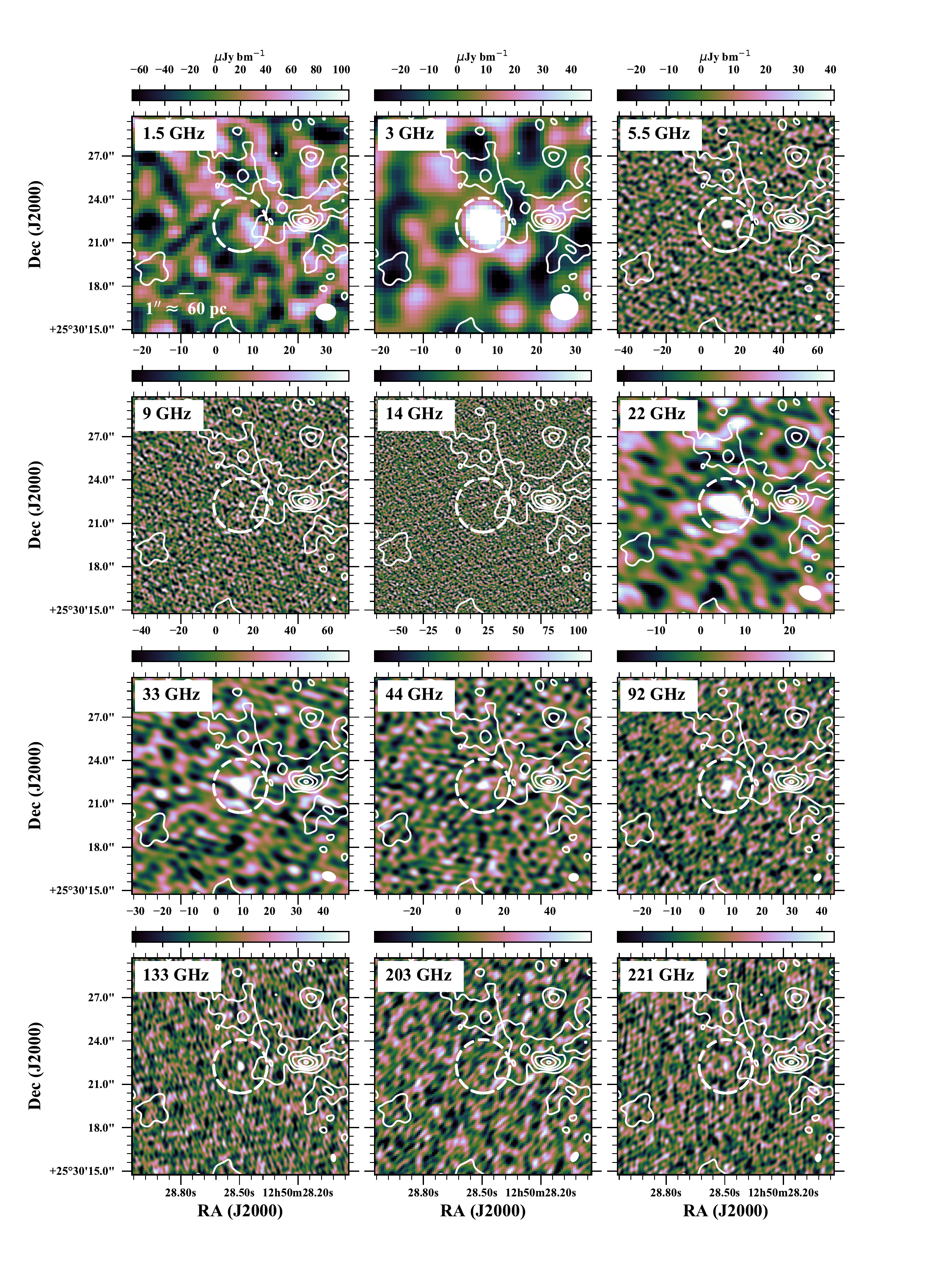}
\caption{Colorscale images displayed with a linear scaling at all observed radio/millimeter frequencies (i.e., 1.5, 3, 5.5, 9, 14, 22, 33, 44, 92, 133, 203, and 221\,GHz) at their native resolutions, indicating that NGC\,4725\,B is detected in all bands except at 1.5\,GHz, however the detection is somewhat marginal at 203\,GHz (ALMA Band\,5), being $\approx 3.1\sigma$.
Colorbars are given in units of $\mu$Jy\,bm$^{-1}$ along the top of each panel. 
Overlaid on each panel are the $J=2\rightarrow1$ CO moment 0 contours starting at the 1$\sigma~( = 0.038\,$Jy\,bm$^{-1}$\,km\,s$^{-1}$) rms level, increasing in increments of 2$\sigma$.  
Each panel is $15\arcsec \times 15\arcsec$ ($\approx 900\,{\rm pc} \times 900$\,pc), which is illustrated in Figure \ref{fig:COextent}.
The circle is centered on the source in each panel, and has a diameter of $3\farcs7$ ($\approx 200$\,pc), which was used for the elliptical Gaussian fitting photometry.  
The size and orientation of the synthesized beams are illustrated in the bottom right corner of each panel.  
A scale bar is shown in the bottom left corner of the first panel.  
}
\label{fig:radpan}
\end{figure*}

\subsection{VLA Data Reduction and Imaging}
Radio data were obtained as part of multiple campaigns using the VLA covering the S- ($2-4$\,GHz), Ku- ($12-18$\,GHz),  K- ($18-26.5$\,GHz), Ka- ($26.5-40$\,GHz, and Q- ($40-50$\,GHz) bands, which are summarized in Table \ref{tbl-1}.  
Details on the data reduction and imaging procedures can be found in \citet{ejm18a,ejm18c,stl20}.  
To summarize, we followed standard calibration procedures, using the VLA calibration pipeline built on the Common Astronomy Software Applications \citep[CASA;][]{casa} versions 4.6.0 and 4.7.0.
After each initial pipeline run, we manually inspected the calibration tables and visibilities for signs of instrumental problems (e.g., bad deformatters) and RFI, flagging correspondingly. After flagging, we re-ran the pipeline, and repeated this process until we could not detect any further signs of bad data.

As with the data calibration, a detailed description of the imaging procedure used here can be found in \citet{ejm18a, stl20}.  
To summarize, calibrated VLA measurement sets for each source were imaged using the task \textsc{tclean} in CASA version 4.7.0.  
The synthesized beam, point source rms, and surface brightness rms for each final image is given in Table~\ref{tbl-1}. 
Employing the ALMA Cycle 4 handbook definition of the largest angular scale (LAS), we find that the LAS these images are sensitive to are $\theta_{\rm LAS} \approx 31\farcs5, 23\farcs5, 16\farcs9, 11\farcs3$, and 8\farcs44~at 3, 15, 22, 33 and 44\,GHz, respectively, where the calculations have been done with the CASA Analysis Utilities task \textsc{estimateMRS}.

We additionally make use of archival VLA data (VLA/12B-191) that included observations toward the nucleus of NGC\,4725 in the L- ($1-2$\,GHz), C- ($4-8$\,GHz, X- ($8-12$), and Ku-bands.
These data were taken in the A-configuration and made use of the 8-bit samplers.
The L-band data had a maximum instantaneous bandwidth of 1\,GHz with a central frequency of 1.5\,GHz.
The data of the remaining frequency bands had a maximum of 2\,GHz of bandwidth with central frequencies of 5.5, 9, and 14\,GHz for the C-, X-, and Ku-band imaging, respectfully.
These data were reduced using the VLA calibration pipeline built on CASA version 5.3.0.  
The synthesized beam, point source rms, and surface brightness rms for each final image is given in Table~\ref{tbl-1}.
Again, using the task \textsc{estimateMRS}, the LAS that these images are sensitive to are $\theta_{\rm LAS} \approx 19\farcs2, 5\farcs26, 3\farcs19$, and 2\farcs05~at 1.5, 5.5, 9, and 14\,GHz, respectively.

Calibrated data were imaged using the task \textsc{tclean} in CASA version 5.6.1.  
The mode of \textsc{tclean} was set to multi-frequency synthesis \citep[{\sc mfs};][]{mfs1,mfs2}. 
We chose to use {\it Briggs} weighting with \textsc{robust=0.5}.  
Given that the fractional bandwidth is $>10\%$ for each band, we set the variable \textsc{nterms=2}, which allows the cleaning procedure to also model the spectral index variations on the sky. 
For the L-band imaging, we additionally made use of the w-projection algorithm to help deal with bright sources far from the phase center.
Unlike the 14\,GHz data, the 15\,GHz data delivered a non-optimal calibration, which was evident in the data of the phase calibrator source itself. 
The resulting 14\,GHz image created from these archival data is therefore used exclusively in the present analysis.

\subsection{ALMA Data Reduction and Imaging}
ALMA data were obtained during Cycle 6 (Project Code: 2018.1.01826.S) in Bands 3, 4, 5, and 6 and are summarized in Table~\ref{tbl-1}.  
Observations at Bands 3, 4, and 5 were centered at 92, 133, and 203\,GHz, respectively, for dedicated continuum imaging; 
observations at each band used 4 spectral windows with a bandwidth of 1.875\,GHz and 31.250\,MHz spectral channels.  
For Band 6 observations, three spectral windows were set up for dedicated continuum imaging similar to Bands 3, 4, and 5 observations. 
However, the fourth spectral window was tuned to 229.612\,GHz with a bandwidth of 1.875\,GHz and spectral resolution of 1938.477\,kHz (2.531\,km\,s$^{-1}$) to detect $^{12}$CO ($J=2\rightarrow1$) at a rest frequency of 230.5424\,GHz.  
The corresponding continuum image created from the line-free Band 6 data is centered at 221\,GHz.  

Like the VLA data, the ALMA data were reduced and calibrated using CASA following standard procedures as part of the ALMA quality assurance process.  
The ALMA data were reduced using both the ALMA calibration pipeline with CASA version 5.4.0, as well as prepared manual reduction and imaging scripts.  
Calibrated data were imaged using the task \textsc{tclean} in CASA version 5.6.1.  
The mode of \textsc{tclean} was set to multi-frequency synthesis \citep[\textsc{mfs};][]{mfs1,mfs2}. 
We chose to use {\it Briggs} weighting with \textsc{robust=0.5}.  
For Bands 3 and 4, where the fractional bandwidth was $>10\%$, we set the variable \textsc{nterms=2}, which allows the cleaning procedure to also model the spectral index variations on the sky. 
For continuum imaging of Bands 5 and 6, we instead set the variable \textsc{nterms=1}; in the case of Band 6, we imaged the 3 dedicated continuum spectral windows and  the line-free portion of the fourth spectral window.  
In all cases of the continuum imaging, we do not make use of the multiscale clean option \citep{msclean,msmfs}.  
The synthesized beam, point source rms, and surface brightness rms for each final image is given in Table~\ref{tbl-1}.  
Based on the task \textsc{estimateMRS}, the LAS that these images are sensitive to are $\theta_{\rm LAS} \approx 6\farcs64, 5\farcs10, 5\farcs76$, and 4\farcs76~at 92, 133, 203, and 221\,GHz respectively.  
In Figure~\ref{fig:radpan}, $15\arcsec \times 15\arcsec$ ($\approx 900\,{\rm pc} \times 900$\,pc) cutouts centered on the location of NGC\,4725\,B are shown for all radio and millimeter data.  

\begin{figure}[t!]
\centering
\epsscale{1.225}
\plotone{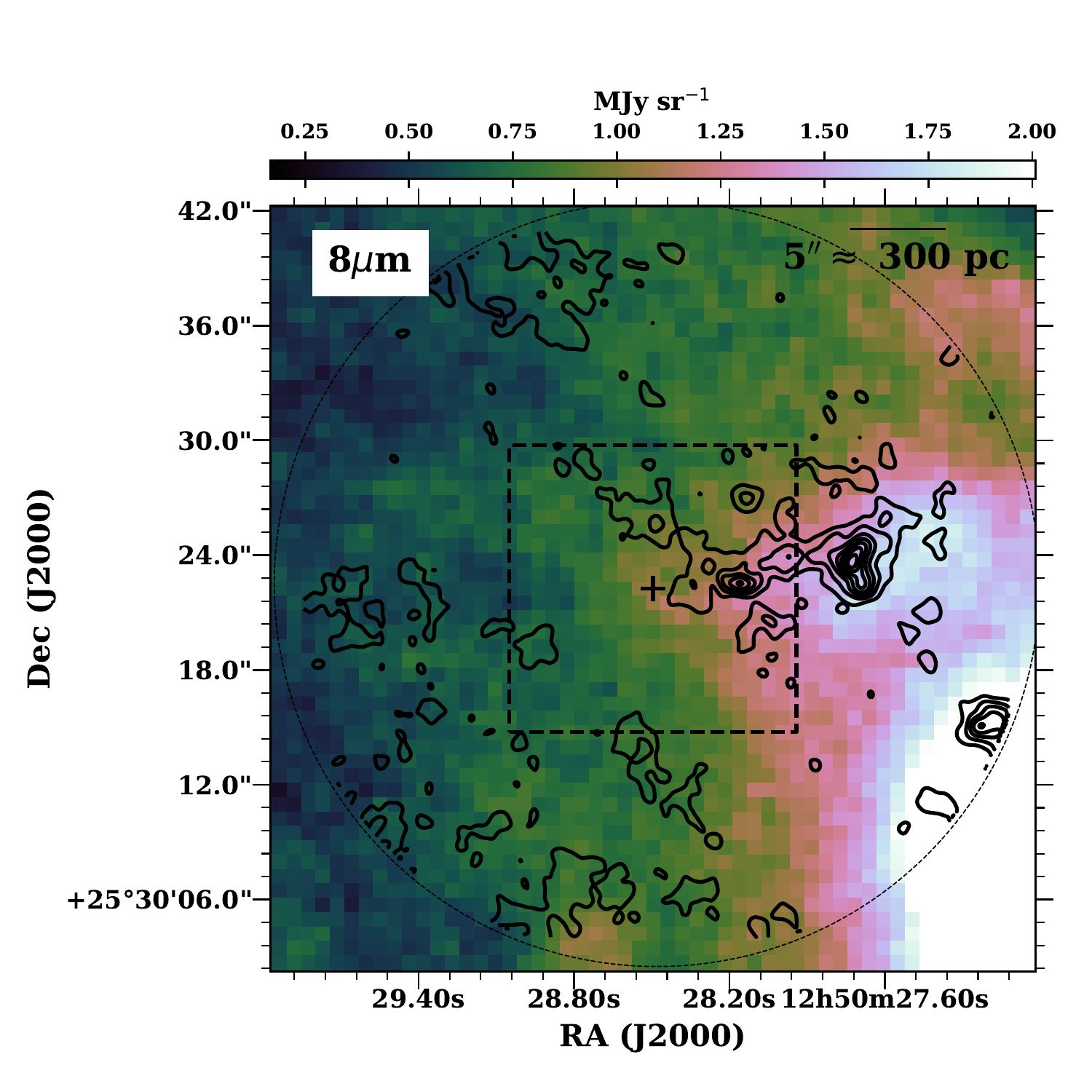}
\caption{A {\it Spitzer}/IRAC 8\,$\mu$m colorscale image of NGC\,4725 ($\approx$2\arcsec~resolution) observed as part of SINGS \citep{rck03}.  
A colorbar is given in units of MJy\,sr$^{-1}$ along the top of the figure. 
The $J=2\rightarrow1$ CO moment 0 contours (at their native resolution) are overlaid starting at the1$\sigma~( = 0.038\,$Jy\,bm$^{-1}$\,km\,s$^{-1}$) rms level, increasing in increments of 2$\sigma$, and illustrating the full extent of our CO imaging; 
the 20\% level of the primary beam response function is shown as a dashed circle.  
The dashed $15\arcsec \times 15\arcsec$ box indicates the size of the panels shown in Figure \ref{fig:radpan}, while the `+' symbol indicates the location of NGC\,4725\,B.  
A scale bar is shown at the top right of the image.  
}
\label{fig:COextent}
\end{figure}

Similar to the continuum imaging, the $J=2\rightarrow1$ CO line data were imaged using \textsc{tclean}, but instead setting the mode to ``cube."    
Unlike the ALMA continuum emission for NGC\,4725\,B, the molecular gas appears extended (see Figure~\ref{fig:COextent}).  
Consequently, we also made use of the multiscale clean option \citep{msclean,msmfs} to help deconvolve extended low-intensity line emission and to search for structures with scales $\approx$1 and 3 times the FWHM of the synthesized beam. 
The moment 0 map was then created by integrating the spectral cube at each pixel limited to the portion of velocity space associated with line emission.  
The synthesized beam and rms of the $J=2\rightarrow1$ CO moment 0 map is given in Table~\ref{tbl-1}.  
The $J=2\rightarrow1$ CO moment 0 contours are overlaid on each radio and millimeter continuum image in Figure~\ref{fig:radpan}.  

\subsection{Photometry}\label{sec:phot}  
Photometry was carried out at all frequencies before applying a primary beam correction.  
To measure the flux densities of the source at each frequency, we first beam-match the data to the lowest resolution beam among all images (i.e., the 3\,GHz data) using the CASA task \textsc{imsmooth}.   
We elect to exclude the much lower resolution 2013 March 33\,GHz data included in \citet{ejm18c}, since it is significantly (i.e., a factor of $\sim$30) coarser than the ALMA data (i.e., $2\farcs89\times1\farcs97$); convolving the data to such a low angular resolution typically reduces/removes the statistical significance of a detection in the higher frequency ALMA bands.  
Further, we make use of the archival VLA 14\,GHz data only rather than the newer 15\,GHz data given the much better calibration and imaging quality of the final map.

Next, the CASA task \textsc{imfit} was used to fit an elliptical Gaussian to the emission within a circular aperture having a radius equal to the FWHM of the synthesized beam major axis of the beam matched images (i.e., 1\farcs85).  
Given the proximity of NGC\,4725\,B to the phase center of each observation, primary beam corrections 
were negligible (i.e., $<1$\%) at all frequencies except 5.5, 9, and 14\,GHz, where correction factors of 1.011, 1.032, 1.079 were applied to the reported peak brightnesses and integrated flux densities (along with their errors), respectively.  
These factors were obtained via the frequency-dependent primary beam correction given in EVLA Memo\#\,195 (Perley 2016)\footnote{\url{http://library.nrao.edu/public/memos/evla/EVLAM_195.pdf}}.    

The source is unresolved in all of the native resolution images using the criterion that the fitted major axis $\phi_{\rm M}$ be at least $2\sigma_{\phi_{\rm M}}$ larger than the FWHM of the synthesized beam major axis.  
Similarly, the source appears unresolved in the beam-matched images at all frequencies. 
Consequently, we take the total flux density ($S_{*}$) for the unresolved cases to be the geometric mean of the peak brightness and integrated flux density reported by \textsc{imfit}.  
We choose this value as it provides the most accurate measurement for the flux density of a weak source given that the uncertainties on the peak brightness and integrated flux density from the elliptical Gaussian fitting are anti-correlated \citep{jc97}.  

The total flux densities, along with their uncertainties, are given in Table~\ref{tbl-1} at each frequency.    
While \textsc{imfit} formally reported a statistically significant detection at 1.5\,GHz (i.e., $S_{*} = 134\pm30$), it is clear from that map (see Figure~\ref{fig:radpan}) that there is nothing co-spatial with NGC\,4725\,B and that a nearby noise feature is being fit.  
Consequently, we consider this as a non-detection and report the 3$\sigma$ rms upper limit in Table~\ref{tbl-1} and exclude this data point from fits.  
There are no significant (i.e., signal-to-noise ratio $> 3$) detections in the beam-matched 203\,GHz continuum image or in the $J=2\rightarrow1$ CO line towards NGC\,4725\,B.  
For the 203\,GHz continuum, we report the \textsc{imfit} results in Table~\ref{tbl-1}, as the fitting function converged and there does appear to be a source at the location of NGC\,4725\,B in the image.  
For the $J=2\rightarrow1$ CO line, \textsc{imfit} failed to converge, so we therefore report the 3$\sigma$ rms upper limit in Table~\ref{tbl-1}.  
The radio-to-millimeter spectrum of NGC\,4725\,B is shown in Figure~\ref{fig:spec}.  


We note that in \citet{ejm18c} they report the integrated 15\,GHz flux density as $S_{*}$ since the source was found to be marginally resolved (i.e., $2.3\sigma$ significance) after convolving all VLA images to the much lower resolution of their 2013 March 33\,GHz.  
Even with differing resolutions and treatment of the source as resolved rather than unresolved, their 15\,GHz flux density is consistent with the 14\,GHz flux density reported in the present study.  

\section{Results}\label{sec:results}
In the following Section, we combine the new ALMA continuum and molecular gas imaging with the existing VLA continuum imaging for NGC\,4725\,B and perform a parametric fit to the data.

\subsection{Parametric Models of the Radio-to-Millimeter Spectrum}
\label{subsec:model}

We employ a simple parametric model to fit the NGC\,4725\,B spectrum presented in Figure~\ref{fig:spec}. The first component in the model is free-free emission, which has flux density

\begin{equation}
    S_\nu^{\rm ff} = B_\nu\left(T_e\right)\left(1-e^{-\tau_\nu^{\rm ff}}\right)\pi\left(\frac{D}{2d_L}\right)^2
    ~~~,
\end{equation}
where $T_e$ is the electron temperature, $B_\nu\left(T\right)$, is the Planck function, $D$ is the diameter of the emitting region, $d_L = 11.9$\,Mpc is the distance to NGC\,4725, and $\tau_\nu^{\rm ff}$ is the free-free optical depth. We model the frequency dependence of $\tau_\nu^{\rm ff}$ in the low-frequency limit ($h\nu \ll kT_e$) following \citet{draine2011}:

\begin{equation}
    \tau_\nu^{\rm ff} = \frac{4}{3}\left(\frac{2\pi}{3}\right)^{1/2} \frac{q_e^6 g_{\rm ff} EM}{\left(m_e k T_e\right)^{3/2}c\nu^2}
    ~~~,
\end{equation}
where $q_e$, $m_e$, $k$, $c$ are, respectively, the electron charge (note Gaussian units), electron mass, Boltzmann constant, and speed of light, and $EM$ is the emission measure. We approximate the Gaunt factor $g_{\rm ff}$ as an analytic function of frequency and electron temperature following \citet{draine2011}:

\begin{equation}
    g_{\rm ff}\left(\nu, T_e\right) = {\rm ln}\Bigg\{{\rm exp} \left[5.960 - \frac{\sqrt{3}}{\pi}{\rm ln}\left(\nu_9 T_4^{-3/2}\right)\right] + e\Bigg\}
    ~~~,
\end{equation}
where $\nu_9 \equiv \nu/10^9\,$Hz and $T_4 \equiv T_e/10^4\,$K. Given the limited number of data points and the relative insensitivity of the spectrum to $T_e$, we fix $T_e = 10^4$\,K. It should be noted, however, that the fit value of $D$ is sensitive to our choice of $T_e$, with higher assumed $T_e$ resulting in smaller fit $D$, and vice-versa.

The spectral slope between 5 and 15\,GHz suggests the presence of an optically-thin synchrotron component, while the sharp decline in emission toward lower frequencies is indicative of free-free absorption. We model the optically-thin synchrotron emission as a power law and the absorption as pure free-free absorption, yielding a spectrum

\begin{equation}
    S_\nu^{\rm sync} = A_s \left(\frac{\nu}{20\,{\rm GHz}}\right)^\alpha e^{-\tau_\nu^{\rm ff}}
    ~~~,
\end{equation}
where $A_s$ is the 20\,GHz flux density in the optically thin limit. We find that the synchrotron component dominates the emission only over a narrow range of frequencies, providing little constraining power on the spectral index. We therefore adopt $\alpha = -0.83$, consistent with similar observations for star-forming galaxies \citep[e.g.,][]{nkw97,ejm11b}.

While the AME component could in principle be modeled using the detailed physical prescription implemented in the \texttt{SpDust} software \citep{ahd09, sah11}, we pursue instead a simple functional description. \citet{cepeda2020} found that the canonical spinning dust spectra in the representative Galactic environments defined by \citet{dl98b} were well-described by

\begin{equation}
\label{eq:ame_sed}
    S_\nu^{\rm AME} = A_{\rm AME}\ {\rm exp}\Big\{-\frac{1}{2}\left[\frac{{\rm ln}\left(\nu/\nu_p\right)}{W_{\rm AME}}\right]^2 - \tau_\nu^{\rm ff}\Big\}
    ~~~,
\end{equation}
in general agreement with the more detailed parameterization derived by \citet{stevenson2014}. Note that, for completeness, we have included a free-free absorption term, even though the effects are small at frequencies where the AME dominates. Here $\nu_p$ is the peak frequency, $A_{\rm AME}$ is the flux density at $\nu_p$, and $W_{\rm AME}$ governs the width of the spectrum. \citet{cepeda2020} found that both the theoretical spectra generated by \texttt{SpDust} and the radio observations of $\lambda$ Ori yielded $W_{\rm AME}$ in the range of 0.4 to 0.7. We caution that this function provides a convenient representation of the spinning dust spectrum near its peak, but is unlikely to accurately represent the low or high frequency tails.

Finally, the high-frequency thermal dust emission is parameterized by a modified blackbody, i.e.,

\begin{equation}
    S_\nu^d = A_d \left(\frac{\nu}{353\,{\rm GHz}}\right)^\beta \frac{B_\nu\left(T_d\right)e^{-\tau_\nu^{\rm ff}}}{B_{353\,{\rm GHz}}\left(T_d\right)}
\end{equation}
where $A_d$ is the flux density at 353\,GHz, $T_d$ is the dust temperature, and $B_\nu\left(T\right)$ is the Planck function. Given the lack of high-frequency data at the requisite angular resolution and the apparent lack of strong signal in the $\nu > 100$\,GHz data, we fix $\beta = 1.55$ and $T_d = 19.6$\,K, which are representative values for dust emission in the Galaxy \citep{Planck_Int_XXII,Planck_2018_XI}. As with AME, the free-free absorption term is added only for completeness and has little effect on the spectrum at millimeter wavelengths.

\subsection{Model Fits}
\label{subsec:fits}

\begin{figure*}[t!]
\epsscale{1.18}
\includegraphics[width=\columnwidth]{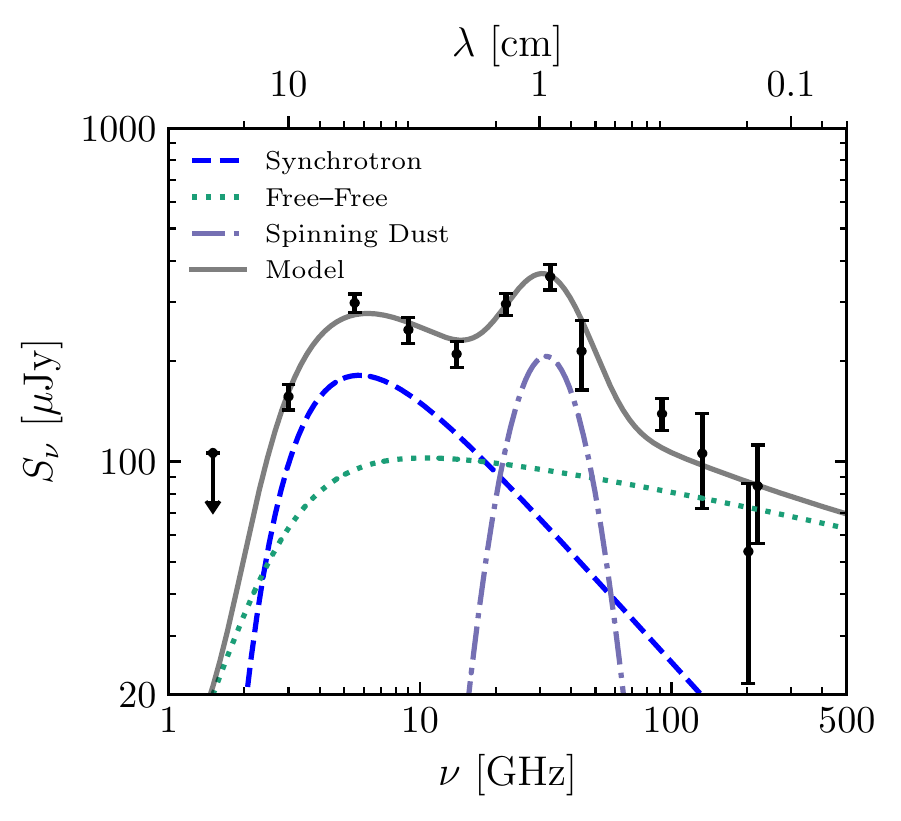}
\includegraphics[width=\columnwidth]{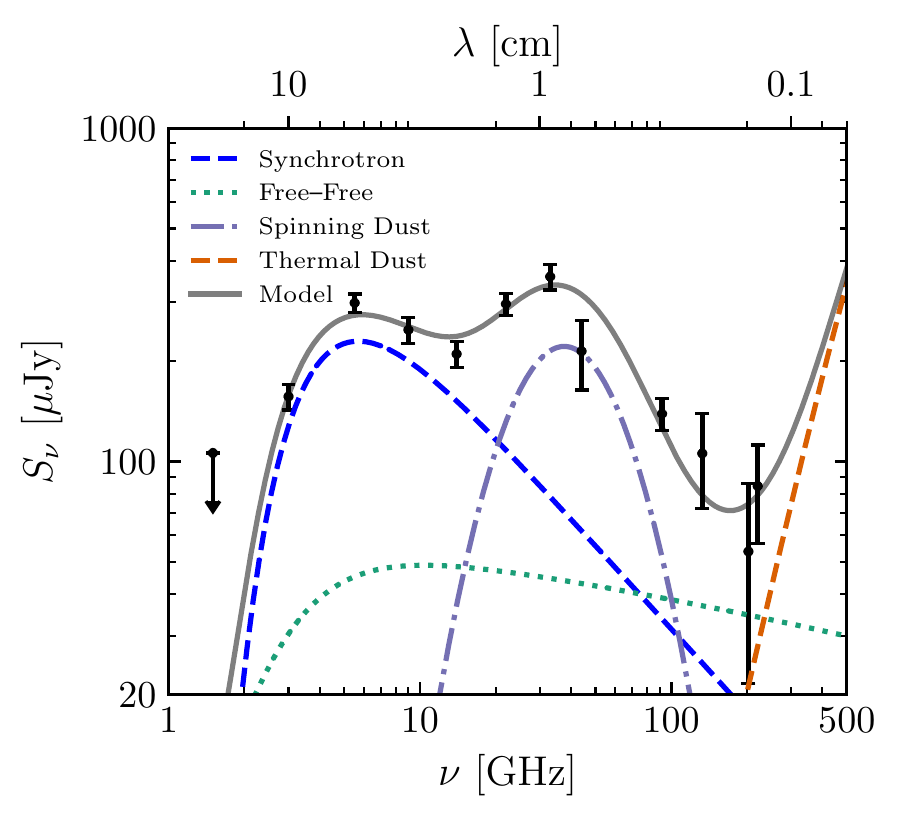}
\caption{The radio-to-millimeter spectrum of NGC\,4725\,B displaying a strong peak at $\approx30$\,GHz along with two best fit models. The model in the left panel includes free-free, spinning dust, and synchrotron emission. The free-free becomes optically thick at 3.7\,GHz. In the right panel, we add millimeter dust emission as a modified blackbody with a temperature of 19.6\,K and $\beta = 1.55$. Both fits provide an adequate description of the data. However, the addition of the dust component results in bimodal posteriors corresponding to the two panels here (i.e., with and without millimeter dust emission). In both fits we find that the free-free emission is consistent with a region having $EM \approx 1.4\times10^{26}\,$cm$^{-5}$ and physical size $D \approx 2$\,pc.}
\label{fig:spec}
\end{figure*}

To fit the model described in the previous section to the observed flux densities, we employ the \texttt{emcee} Markov Chain Monte Carlo software \citep{emcee} to sample the seven parameters $EM$, $D$, $A_s$, $A_{\rm AME}$, $W_{\rm AME}$, $\nu_p$, and $A_d$. We use broad, uninformative priors, requiring only that $EM$, $D$, $A_s$, $A_{\rm AME}$, and $A_d$ be positive, that $10 < \nu_p/{\rm GHz} < 50$, and that $0 < W_{\rm AME} < 1$.

The full seven parameter fit yields bimodal posteriors corresponding to two distinct scenarios, which we illustrate in Figure~\ref{fig:spec}. In the first scenario (Figure~\ref{fig:spec}, left panel), the millimeter dust emission is consistent with zero and the AME is sharply peaked near 30\,GHz. In the second scenario (Figure~\ref{fig:spec}, right panel), thermal dust and free-free emission contribute roughly equally to the emission in the ALMA bands, the AME component is significantly broader and shifted to higher frequencies ($\nu_p = 38$\,GHz), and the synchrotron emission dominates the low frequency spectrum. The model without millimeter dust emission is the maximum likelihood solution, and so we consider this our default model and quote parameter constraints from a six parameter fit fixing $A_d = 0$. Inclusion of a small amount of dust emission does not qualitatively alter this fit and cannot be ruled out; this model should be considered as representative of the scenario in which emission in the ALMA bands is mostly free-free. As we discuss in greater detail in Section~\ref{subsub:ame_constraints}, the alternate scenario with significant thermal dust emission at millimeter wavelengths does not modify our principal conclusions, including the fact that the millimeter emission is much weaker than anticipated if the 30\,GHz excess is attributed to spinning dust. The full posterior distributions are presented in Figure~\ref{fig:triangle} in the Appendix.

Despite the simplicity of the model with no millimeter dust emission (i.e., $A_d = 0$), it provides an excellent description of the data ($\chi^2 = 9.0$ for a six parameter fit to eleven data points). To match the turnover at low frequencies, free-free absorption is required. We find that the free-free emission can be explained by a region with $EM = \left(1.46\pm0.15\right)\times10^{26}\,$cm$^{-5}$ and physical size $D = 2.30^{+0.30}_{-0.49}$\,pc. The free-free optical depth is unity at 3.7\,GHz. In general agreement with the analysis of \citet{ejm18c}, we find that free-free accounts for only $25\%$ of the 30\,GHz emission, with AME dominating at this frequency. We differ in our analysis by the inclusion of synchrotron emission, which contributes 18\% of the 30\,GHz emission in this model. The 30\,GHz excess is well fit with an AME component described by Equation~\ref{eq:ame_sed} and having peak frequency of $\nu_p = 31.6^{+6.5}_{-3.0}\,$GHz and a rather narrow width of $W_{\rm AME} = 0.32^{+0.20}_{-0.11}$. For completeness, the remaining model parameters have fit values of $A_s = 95^{+21}_{-17}$\,$\mu$Jy and $A_{\rm AME} = 209^{+47}_{-37}\,\mu$Jy.

The greatest discrepancy between the data and the model is at 92\,GHz, where the model is 1.9$\sigma$ too low, but this may also be an indication that the simple functional form used to fit the AME component (Equation~\ref{eq:ame_sed}) falls off too rapidly at high frequencies. More observations are needed to better constrain the shape of the spectrum, particularly at frequencies above 30\,GHz.

The alternative solution with nonzero millimeter dust emission differs significantly in the relative importance of the emission mechanisms. The observed 30\,GHz emission is accounted for by 202\,$\mu$Jy of AME (62\% of the total), 87\,$\mu$Jy of synchrotron (27\%), and only 37\,$\mu$Jy of free-free (11\%) in this model. The greatly reduced free-free emission allows for the incorporation of dust emission at higher frequencies. In this scenario, the millimeter dust continuum and free-free emission contribute equally at 228\,GHz. The 95\% upper limit on dust emission at 353\,GHz is 167\,$\mu$Jy. The AME component is much broader ($W_{\rm AME} = 0.5$) and the peak frequency is shifted to 38\,GHz. However, the quality of the fit is poorer even with the addition of the extra fitting parameter ($\chi^2 = 10.3$ fitting seven parameters to eleven data points). On the other hand, it would be surprising for NGC\,4725B to lack millimeter dust emission entirely.

\subsection{Associated Molecular Gas: $^{12}$CO ($J=2\rightarrow1$)}
\label{subsec:assoc_gas}
In Figure \ref{fig:COextent}, the $J=2\rightarrow1$ CO moment 0 contours are plotted on a {\it Spitzer}/IRAC 8\,$\mu$m colorscale image extending out to the 20\% level of the primary beam response, indicating a significant amount of molecular gas present in this vicinity of the star-forming disk. This is consistent with expectations based on existing lower resolution (i.e., $\theta_{1/2} \approx 13\farcs4$) $J=2\rightarrow1$ CO  data \citep{heracles09}.  
The molecular gas distribution loosely follows the 8\,$\mu$m brightness, which is a tracer of small dust grains and polycyclic aromatic hydrocarbons (PAHs).  
Figure \ref{fig:radpan} shows a zoom-in of the $J=2\rightarrow1$ CO moment 0 contours overlaid on each of the radio and millimeter colorscale images.  

The fact that the CO ($J=2\rightarrow1$) does not peak exactly at the location of continuum emission from NGC\,4725\,B on $\lesssim100$\,pc scales is not surprising \citep[e.g.,][]{ejm15} given that the molecular gas has already been converted into stars, with plenty of additional gas nearby available for future star formation. 
The integrated line luminosity of the nearest CO feature $\approx260$\,pc to the west of NGC\,4725\,B is $L_{\rm CO}^{\prime}\approx (2.2\pm0.23)\times10^{5}$\,K\,km\,s$^{-1}$\,pc$^{2}$. Assuming both a CO ($J=2\rightarrow1$)/CO ($J=1\rightarrow0$) brightness temperature ratio of $\sim$0.8 \citep[e.g.,][]{heracles09,lzk18} and $\alpha_{\rm CO} \approx 4.3$\,$M_{\odot}\,{\rm (K\,km\,s^{-1}\,pc^{2})}^{-1}$ \citep[e.g.,][]{svb05,bwl13}, the total nearby molecular gas reservoir is measured to be $M_{\rm H2}\gtrsim 1.2\times10^{6}\,M_{\odot}$.
For comparison, given the 3$\sigma$ upper-limit of $\lesssim 7.1 \times 10^{4}$\,K\,km\,s$^{-1}$\,pc$^{2}$ at the location of NGC\,4725\,B, the corresponding upper limit on the molecular gas mass is $M_{\rm H_{2}}\lesssim 3.8\times10^5\,M_{\odot}$.

\section{Discussion and Conclusions}\label{sec:summary}
We discuss here the possible origins of the emission observed from NGC\,4725\,B, including spinning dust emission in a massive cluster and synchrotron emission in a background radio source. Neither provides a completely satisfactory explanation of the spectrum and the apparent compactness.

\subsection{A Forming Massive Cluster?}
\subsubsection{Cluster Properties}

The new ALMA data presented in this study appear consistent with the conclusion of \citet{ejm18c} that NGC\,4725\,B is a nascent star-forming region, where massive stars are still highly enshrouded by their natal cocoons of gas and dust. 
The archival VLA data suggest the presence of synchrotron emission in addition to a significant amount of free-free emission, and thus an age of $\sim 3-5$\,Myr. 
Assuming that $\approx 50$\% of the 14\,GHz flux density arises from free-free emission, based on the best-fit model in the left panel of Figure~\ref{fig:spec}, suggests an ionizing photon rate of $Q(H^{0}) \approx 1.4\times10^{51}\,$s$^{-1}$ \citep[][Equation 10]{ejm11b}, or $\approx 125$ O7V stars, where $Q(H^{0}) \approx 10^{49}\,$s$^{-1}$ per 37.7\,$M_{\odot}$ O7V star \citep{sternberg03}. 
Using a Kroupa initial mass function \citep{pk01} fully populated between 0.1 and 100\,$M_{\odot}$ implies a total stellar mass of $M_{\star} \approx 6.4\times10^{4}\,M_{\odot}$ for NGC\,4725\,B. 
It is worth stressing that our results and conclusions remain qualitatively the same even if we were to instead use the best-fit model in the right panel of Figure~\ref{fig:spec}. 
In that case, the fraction of free-free emission at 14\,GHz reduces to $\approx$20\% of the total flux density, which in turn reduces the associated ionizing photon rate and stellar mass estimate by $\approx40\%$ [i.e., $Q(H^{0}) \approx 5.9 \times 10^{50}\,$s$^{-1}$ and $M_{\star} \approx 2.6\times10^{4}\,M_{\odot}$].  

With a stellar mass of $\lesssim 10^{5}\,M_{\odot}$ located in a $\sim 2$\,pc region, NGC\,4725\,B is best defined as a young massive star cluster \citep[YMC;][]{pz10}. In that context, we find that the mass NGC\,4725\,B is consistent with the upper-end of star cluster masses derived for other nearby galaxies (e.g., NGC\,628 and M\,51) as part of the Legacy Extragalactic UV Survey \citep[LEGUS;][]{dc15}. Further, the estimated cluster radius of $\sim 1$pc agrees with observations of the youngest YMCs studied in the LEGUS sample \citep{ryon17}.
Finally, radial gradients in the maximum cluster mass have been observed for several nearby spiral galaxies, such that clusters with $M_{\star} \gtrsim 10^{4}\,M_{\odot}$ are found predominately within the central $\sim 2$\,kpc \citep{aa17,mm18a}. This is consistent with NGC\,4725\,B which is located at $\sim 1.9$\,kpc from the nucleus of NGC\,4725.

Using a combination of ALMA, {\it HST}, and VLA observations of the Antennae Galaxies, \citet{bcm14} developed a classification system for YMCs such that the presence and strength of different multiwavelength tracers of star formation activity can be used as an independent metric to age date star clusters. 
With bright and compact radio emission, a large inferred value for the dust attenuation, and the presence of nearby diffuse/faint CO (see Section~\ref{sec:molgas}), NGC\,4725\,B is classified as a stage 3 ``emerging cluster," which places an upper-limit on the inferred cluster age of $\lesssim 3-5$\,Myr.  
This is consistent with the age inferred from the radio emission alone.  
  
During this phase of cluster evolution, the natal gas and dust surrounding the cluster is actively removed via feedback from stellar winds and UV radiation \citep{ll03}. In this context, the lack of CO gas and millimeter dust emission observed is perhaps not that surprising or unexpected. 
Furthermore, while there is not a significant detection at the exact location of NGC\,4725\,B, the molecular gas appears to be in the immediate vicinity of NGC\,4725\,B (i.e., $
\lesssim100$\,pc) and could have fueled the nascent star formation in NGC\,4725\,B.  
Assuming an age of $\sim 5$\,Myr for the star formation and a distance of $\sim$100\,pc to the nearby molecular gas reservoir suggests an asymmetric drift velocity of $\sim$20\,km\,s$^{-1}$ between NGC\,4725\,B and the CO gas, which seems completely plausible \citep[e.g.,][]{Quirk19}.

\citet{miura12} used CO observations of M\,33 to classify giant molecular clouds (GMCs) into different evolutionary stages.  Given the number of O7V stars estimated above, NGC\,4725\,B and the associated GMC would be classified as a stage ``C" star-forming region with an age of $\sim 3-5$\,Myr and a mean CO ($J=2\rightarrow1$) flux of $L_{\rm CO}^{\prime} \sim 7.2 \times 10^{4}$\,K\,km\,s$^{-1}$\,pc$^{2}$ with an associated standard deviation of $\sigma_{L^{\prime}_{\rm CO}} \sim 4.6\times 10^{4}$\,K\,km\,s$^{-1}$\,pc$^{2}$.  
Given the large dispersion, this value is consistent with our non-detection (i.e., $L_{\rm CO}^{\prime} \lesssim 7.1 \times 10^{4}$\,K\,km\,s$^{-1}$\,pc$^{2}$) at the location of NGC\,4725\,B.  

\subsubsection{Molecular Gas Properties \label{sec:molgas}}

Let us assume that NGC\,4725\,B formed from a cloud of similar mass in the nearby molecular gas reservoir to the west (i.e., $M_{\rm H2} \approx 1.2\times10^{6}\,M_{\odot}$; see Section~\ref{subsec:assoc_gas}) .  
Using \citet[][Equation 1]{ejm11b}, we convert the value of $Q(H^{0})$ given above to a star formation rate of 0.010\,$M_\odot$\,yr$^{-1}$ for NGC\,4725\,B. 
This implies a gas depletion time of $t_{\rm dep} \gtrsim 113$\,Myr.  
The gas depletion time inferred here is a factor of $\approx$4 shorter than that for starburst galaxies in the Great Observatories All-Sky LIRG Survey \citep[GOALS;][]{lee09,rhi19} and a factor of $\approx$16 shorter than that derived for resolved measurements on $60-120$\,pc scales for normal star-forming galaxies included the Physics at High Angular Resolution in nearby Galaxies (PHANGS)-ALMA sample \citep{utomo18}.

Assuming a constant 100\,pc thickness of the molecular gas layer for this CO complex suggests a gravitational free-fall time of $t_{\rm ff}\sim 6$\,Myr and a star formation efficiency per gravitational free-fall time of $\epsilon_{\rm ff}\sim 5\%$.  
The inferred values of $\tau_{\rm ff}$ and $\epsilon_{\rm ff}$ for the CO feature near NGC\,4725\,B are a factor of $\approx$2 shorter and a factor of $\approx$8 larger than the corresponding average values reported for the PHANGS-ALMA sample \citep{utomo18}, respectively, suggesting relatively efficient star formation.  
This high star formation efficiency per gravitational free-fall time is further supported by taking the ratio of estimated stellar mass in NGC\,4575\,B to the nearby total molecular gas mass, which implies that $\lesssim 5\%$ of the gas has recently been converted into stars. 

Simulations of turbulent, magnetized star-forming clouds (i.e., with a mass of $\sim 10^{5}\,M_{\odot}$) that include stellar radiation and outflow feedback predict a full range of $\epsilon_{\rm ff}$ spanning $\lesssim 1-100\%$. In these simulations $\epsilon_{\rm ff}$ is small at low densities. As mass flows rapidly into higher density regions within the cloud, $\epsilon_{\rm ff}$ approaches unity \citep{khullar19} with little time spent at intermediate densities.  Therefore, the $\epsilon_{\rm ff}$ derived for NGC\,4725\,B may suggest that this object is being observed in a short-lived phase of its evolution. 

In this context, it may not be unexpected that the AME signatures in this region are unique relative to the AME seen in our Galaxy. 
The Galaxy contains no YMCs with masses $\gtrsim 10^{5}\,M_{\odot}$, and if these particular properties of the AME are indeed associated with a short-lived period of massive cluster evolution, we would not expect to observe such interstellar medium conditions in our Galaxy.

\subsubsection{Physical Constraints on AME}
\label{subsub:ame_constraints}

Surprisingly, even with the addition of ALMA data at 92, 133, 203, and 221\,GHz, there is no evidence for high-frequency thermal dust emission at these sensitivities. As discussed in Section~\ref{subsec:model}, we derive an upper limit (95\% confidence) on the 353\,GHz flux density from dust emission of 167\,$\mu$Jy. To place this limit in context, we note that the Galaxy has a typical ratio of 30\,GHz AME specific intensity to 353\,GHz optical depth $\tau_{353}$ of 200\,MJy\,sr$^{-1}$, whether in the diffuse interstellar medium or in cloud complexes like Perseus \citep{bh16,AME18}. For an assumed dust temperature of 20\,K and 1\farcs85 beam, our fit suggests $\tau_{353} < 4\times10^{-5}$. Combining this with a peak AME flux density of 208\,$\mu$Jy yields a ratio of $6\times10^4$\,MJy\,sr$^{-1}$, roughly 300 times larger than typically seen in the Galaxy. The maximum likelihood solution (left panel of Figure~\ref{fig:spec}) invokes no dust emission at all, greatly exacerbating this discrepancy.

Given the surprising lack of millimeter dust emission inferred from the full model fit, we quantify the maximum permitted by the ALMA data without accounting for other emission mechanisms. Fitting just the 203 and 221\,GHz ALMA observations with the modified blackbody dust model, we derive a 95\% upper limit of 520\,$\mu$Jy at 353\,GHz. Even in this maximally conservative scenario, the ratio of AME to millimeter dust emission exceeds Galactic values by a factor of 100.

\subsubsection{Lack of Millimeter Dust Continuum}
Emission from $\sim20\,$K grains is almost certainly present in the region based on much coarser far-infrared data, however it may be present only on much larger spatial scales that are invisible to the ALMA interferometric observations (i.e., $\gtrsim6\arcsec$). 
For instance, there is clearly widespread diffuse {\it Spitzer}/IRAC 8\,$\mu$m emission all around the vicinity of NGC\,4725\,B (see Figure \ref{fig:COextent}), indicating the presence of small dust grains and PAHs. 
As was shown in \citet[][Figure 3]{ejm18c}, there appear to be counterparts at the location of NGC\,4725\,B detected in IRAC maps both at 3.6 and 4.5\,$\mu$m, indicating the presence of small/hot dust grains that could be powering the observed AME. 

We measure a 3.6\,$\mu$m flux density of 974\,$\mu$Jy integrating in a $2\farcs4$ aperture and applying an aperture correction factor of 1.215 as given in Table~4.7 of the IRAC instrument handbook. In the 3.5\,$\mu$m DIRBE band, \citet{Dwek1997} measured a typical dust emission per H-atom of $1.18\times10^{-23}$\,MJy\,sr$^{-1}$\,H$^{-1}$ while \citet{Planck_Int_XVII} measured a value of $6.7\times10^{-24}$\,MJy\,sr$^{-1}$\,H$^{-1}$ at 100\,GHz. If this ratio of mid-infrared to millimeter emission holds for NGC\,4725\,B, we would expect 550\,$\mu$Jy of dust emission at 100\,GHz, well in excess of the total observed emission ($\sim130\,\mu$Jy). The differing resolutions of the {\it Spitzer} and ALMA data and the imperfect correspondence between the IRAC and DIRBE bandpasses may account for some of this discrepancy. However, it is likely that the radiation field is stronger in NGC\,4725B than on a typical diffuse Galactic sightline. A factor of $\sim 5$ difference in the ratio of PAH to submillimeter emission could be accommodated by a factor of $\sim 5$ difference in the strength of the intensity of starlight heating the grains. Thus, the lack of submillimeter continuum at the current sensitivity limits is not implausible given the observed PAH emission in the vicinity of this region.

NGC\,4725B lacks not only dust continuum detectable at these sensitivities, but also any appreciable CO emission. In Section~\ref{subsec:assoc_gas}, we found that the upper limit on $^{12}$CO ($J=2\rightarrow1$) implies $M_{{\rm H}_2} \lesssim 3.8\times10^5\,$M$_\odot$. At 230\,GHz, the \citet{dl07} dust model emits $2.1\times10^{-27}$\,erg\,s$^{-1}$\,H$^{-1}$ when illuminated by starlight characteristic of the diffuse Galactic ISM. Using this emissivity, $4\times10^5\,$M$_\odot$ of gas at a distance of 11.9\,Mpc corresponds to only 26\,$\mu$Jy of continuum emission at 230\,GHz. Thus, the upper limit on the gas mass implied by the non-detection of CO is completely consistent with the lack of observed dust emission.

On the basis of these arguments, the lack of millimeter dust continuum is completely consistent with the observed synchrotron, free-free, CO, and PAH emission in NGC\,4725\,B, supporting the interpretation that this object is a massive cluster within the galaxy disk. However, this scenario does not readily explain the peaked 30\,GHz emission. If it is attributed to spinning dust grains, then the millimeter dust continuum should be roughly two orders of magnitude brighter than observed.

\subsection{A Background Source?}

The compact nature of NGC\,4725 and extreme ratio of 30\,GHz AME relative to 353\,GHz dust emission are clearly at odds with expectations from spinning dust emission.
However, as discussed in \citet{ejm18c}, attributing NGC\,4725\,B to a chance background galaxy is also problematic, though cannot be completely ruled out.  
First, there are no known submillijansky radio sources that peak at such high frequencies. 
And, while the submillijansky extragalacitc sky at $\gtrsim 30$\,GHz is not well known, naively extrapolating from existing number counts \citep[e.g., ][]{massardi11, ejm18b} suggest that finding a source with a rising spectrum between 5 and 20\,GHz given the SFRS survey area is unlikely \citep[i.e., $<17$\%,][]{ejm18c}.
This is a conservative upper limit given that this estimate both assumes uniform noise in the SFRS survey area and uses the full sample of \citet{massardi11}, which also finds that the fraction of sources with rising spectra decreases with decreasing flux density.  
For example, applying an average primary beam correction to the SFRS noise and extrapolating from the fraction of sources with rising spectra between 5 and 20\,GHz in their lowest flux density bin (i.e., $40 \leq S_{\rm 20\,GHz} < 100$\,mJy, which is still more than 100$\times$ brighter than NGC\,4725\,B) decreases the likelihood of finding such a source to $<10$\%.  
Assuming this trend continues, this would then suggest an even lower likelihood of finding such a source given the 22\,GHz flux density of NGC\,4725\,B (i.e., $S_{\rm 22\,GHz} \approx 300\,\mu$Jy).    

The spectrum of NGC\,4725\,B also seems much too sharply peaked to be explained by a ``classical" (synchrotron-self absorbed) gigahertz-peaked spectrum (GPS) radio galaxy \citep{ERA}.  
Furthermore, if the source were a GPS source at redshift $z\approx1$ with a turnover frequency of 31.6\,GHz (i.e., 63.2\,GHz in the source frame) and a corresponding flux density of $\sim365\,\mu$Jy, the empirical turnover-size relation given by \citet{Orienti14} suggests a physical diameter of $D \approx 0.4$\,pc.  
This would then imply an internal magnetic field strength of $\sim6\times10^{5}$\,G \citep{ERA}, which is extraordinarily large.  
Assuming a spherical source, the corresponding magnetic energy [$E_{B} = U_{B}V$, where $U_{B} = B^2/(8\pi)$ is the magnetic energy density and $V$ is the source volume] would be $\sim 1.1 \times 10^{64}$\,erg, exceeding the gravitational binding energy ($E_{G} \approx 2GM^2/D$) unless the enclosed mass was $M > 1.6\times10^{11}\,M_{\odot}$!  
While this empirical turnover-size relation was derived for sources with typical flux densities as large as $\sim100\,$mJy (i.e., $\sim$300$\times$ larger than NGC\,4725\,B), by locating it at $z\approx1$, its rest-frame 5\,GHz spectral luminosity would be consistent with the low luminosity end of the core emission in that sample (i.e., $L_{\rm 5\,GHz} \sim 1.1\times 10^{24}$\,W\,Hz$^{-1}$).  
Thus, such a scenario seems implausible.

Although it remains possible that NGC\,4725\,B is a new class of compact, optically-faint, $\sim30$\,GHz peaking radio galaxy, on the basis of these arguments it still seems most likely that it is located in the disk of NGC\,4725. 
Future high resolution (i.e., $\lesssim 0\farcs01$) radio imaging may be able to discriminate between these scenarios.

\subsection{Final Thoughts}
While the expectation was that these new ALMA data would help definitively characterize the physical conditions of NGC\,4725\,B and explain why there is such a strong presence of AME, they have in fact raised further questions. It is extraordinary that NGC\,4725\,B has strongly peaked emission at $\approx1$\,cm without any detectable dust emission at millimeter wavelengths.

If the excess 30\,GHz emission is being produced by spinning ultrasmall grains, the extreme ratio of 30\,GHz AME relative to 353\,GHz dust emission suggests, at minimum, substantial differences in the small grain population in this region with respect to those in Galactic AME regions. Models of Galactic dust that reproduce both far-infrared emission and AME require at least $\sim$5\% of the dust mass to reside in grains $\lesssim1$\,nm in size \citep{dl98b,Hoang16,hd17}. Thus, a factor of $\sim$300 enhancement in the 30\,GHz emissivity relative to diffuse Galactic sightlines could not be achieved even if 100\% of the dust mass were in nanoparticles. Other factors would be required, such as grains having much larger typical electric dipole moments while maintaining comparable rotation frequencies. Another possibility might be a scenario in which all of the grains were reduced to a sufficiently small size ($\lesssim 5\,$nm) as not to emit efficiently at long wavelengths. While not completely implausible, these considerations cast doubt on a spinning dust origin of the 30\,GHz emission.

Alternatively, an emission mechanism such as thermal magnetic dipole emission \citep{dl99,dh13} might produce elevated 30\,GHz relative to 353\,GHz emission through systematic changes in the physical properties of the grains. However, such an explanation would require significant fine-tuning of the dust optical properties and aspect ratios to achieve such strong ferromagnetic resonance behavior, and we thus consider it unlikely.

At this point, it seems that more observations are required to make further progress on determining the true physical nature of this source and why there appears to be such strong AME.
For instance, significantly deeper, high angular resolution VLA/VLBA and ALMA data should help to better constrain the spectrum (e.g., determine if the ALMA measurements are completely dominated by  pure free-free emission) and angular extent of NGC\,4725\,B.  
Looking toward the future, next-generation ground- and space-based facilities should contribute substantially to such studies.  
A next-generation VLA \citep[ngVLA;][]{ngvla-sci, ngvla-astro2020}, delivering an order of magnitude improvement and in sensitivity and angular resolution than the existing VLA, will be highly optimized in studying faint sources like NGC\,4725\,B at frequencies spanning $\sim1.2-116$\,GHz.  
While future far-infrared measurements at a high enough angular resolution to constrain the total infrared emission from NGC\,4725\,B will need to wait a decade or more for a space-based telescope, such as the currently conceived {\it Origins Space Telescope}, observing the 3.3 and 3.4\,$\mu$m hydrocarbon features with the Near-Infrared Spectrograph on board the {\it James Webb Space Telescope} could help shed critically needed light on potential AME carriers. 

\acknowledgements
We would like to thank the anonymous referee for comments that helped to improve the content and presentation of this letter.
E.J.M thanks J.J. Condon for useful discussions that helped improve the paper.  
The National Radio Astronomy Observatory is a facility of the National Science Foundation operated under cooperative agreement by Associated Universities, Inc. 
This paper makes use of the following ALMA data: ADS/JAO.ALMA\#2018.1.01826.S. ALMA is a partnership of ESO (representing its member states), NSF (USA) and NINS (Japan), together with NRC (Canada), MOST and ASIAA (Taiwan), and KASI (Republic of Korea), in cooperation with the Republic of Chile. The Joint ALMA Observatory is operated by ESO, AUI/NRAO and NAOJ.
\software{APLpy \citep{aplpy2012,aplpy2019}, CASA \citep[v4.6.0, v4.7.0, v5.3.0, v5.4.0, v5.6.1][]{casa}, \texttt{corner} \citep{corner}, \texttt{emcee} \citep{emcee}, Matplotlib \citep{Matplotlib}, NumPy \citep{NumPy}}

\bibliography{master_ref}

\begin{figure*}[t!]
\epsscale{1.18}
\includegraphics[width=\textwidth]{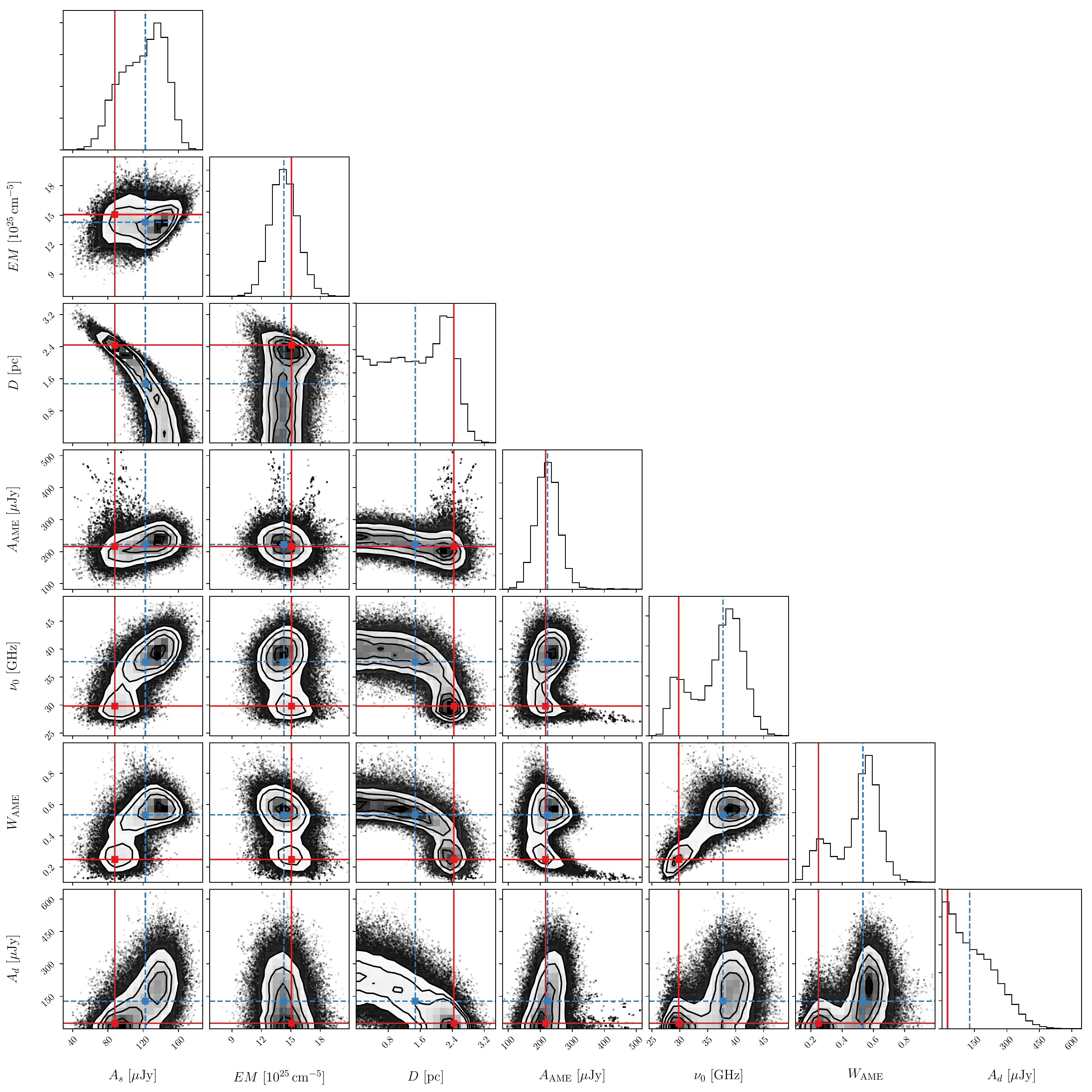}
\caption{Posterior distributions for all seven fit parameters of the model described in Section~\ref{subsec:model}. The maximum likelihood values are indicated with the red solid line and correspond to the solution with no millimeter dust emission (i.e., $A_d = 0$). The posterior medians are indicated with the blue dashed lines and correspond roughly to the other local maximum in the posterior distributions.}
\label{fig:triangle}
\end{figure*}

\appendix
\label{sec:appendix}
In Figure~\ref{fig:triangle}, we present the full posteriors for all seven parameters in the model described in Section~\ref{subsec:model}.

\end{document}

%% file: obsdata.tex
\begin{deluxetable}{c|cccc}
\tablecaption{Observational Data \label{tbl-1}}
\tabletypesize{\scriptsize}
\tablewidth{0pt}
\tablecolumns{5}
\tablehead{
\colhead{Freq.} & \colhead{Synthesized}& \colhead{$\sigma$}& \colhead{$\sigma_{T_{b}}$}& \colhead{$S_{*}^{\dagger}$}\\
\colhead{(GHz)} & \colhead{Beam}& \colhead{($\mu$Jy\,bm$^{-1}$)}& \colhead{(mK)}& \colhead{($\mu$Jy)}
}
\startdata
    1.50  &  $1\farcs32 \times 1\farcs09$  & 31.1  &  11697.  &         $<$  106.  \\
    3.00  &  $1\farcs85 \times 1\farcs76$  & 13.7  &  569.37  & $  157.\pm   14.$  \\
    5.50  &  $0\farcs36 \times 0\farcs31$  & 12.2  &  4387.4  & $  299.\pm   19.$  \\
    9.00  &  $0\farcs25 \times 0\farcs21$  & 10.8  &  3125.6  & $  248.\pm   22.$  \\
    14.0  &  $0\farcs15 \times 0\farcs12$  & 10.1  &  3478.4  & $  210.\pm   19.$  \\
    22.0  &  $1\farcs50 \times 0\farcs89$  & 20.5  &  38.340  & $  297.\pm   23.$  \\
    33.0  &  $0\farcs91 \times 0\farcs58$  & 21.4  &  45.310  & $  359.\pm   31.$  \\
    44.0  &  $0\farcs63 \times 0\farcs50$  & 32.9  &  66.700  & $  214.\pm   51.$  \\
    92.0  &  $0\farcs54 \times 0\farcs37$  & 8.88  &  7.4800  & $  139.\pm   15.$  \\
    133.  &  $0\farcs48 \times 0\farcs29$  & 14.8  &  8.1000  & $  106.\pm   33.$  \\
    203.  &  $0\farcs69 \times 0\farcs40$  & 17.2  &  1.9800  & $  53.7\pm   32.$  \\
    221.  &  $0\farcs51 \times 0\farcs34$  & 13.4  &  1.8000  & $  84.3\pm   28.$  \\
\cutinhead{$^{12}$CO ($J=2\rightarrow1$) Emission Line}
 & & (Jy\,bm$^{-1}$\,km\,s$^{-1}$)&(K\,km\,s$^{-1}$) & (Jy\,km\,s$^{-1}$) \\
\hline
229.6  &  $0\farcs61 \times 0\farcs46$  &0.038  & 3.15  &         $<$ 0.820    
\enddata
\tablenotetext{\dagger}{Photometry measured after convolving all data to match the synthesized beam of the 3\,GHz data.}
\end{deluxetable}